\documentclass[prl,twocolumn,showpacs]{revtex4}
\usepackage{amsmath,amssymb}

\begin{document}
\title{Response to ``CPT symmetry and antimatter gravity in general relativity''}
\author{Daniel J. Cross}
\affiliation{Bryn Mawr College}
\begin{abstract}
The observed accelerated cosmic expansion is problematic in that it seems to require an otherwise unobserved dark energy for its origin.  A possible alternative explanation has been recently given, which attempts to account for this expansion in terms of a hypothesized matter-anti-matter repulsion.  This repulsion or anti-gravity is derived by applying the CPT theorem to general relativity.  We show that this proposal cannot work for two reasons: 1) it incorrectly predicts the behavior of photons and 2) the CPT transformation itself is not consistently applied.
\end{abstract}

\maketitle

In a recent letter, Villata \cite{V2011} proposed a novel mechanism to drive the accelerated cosmic expansion: a mutual gravitational repulsion of matter and anti-matter.  By then locating anti-matter within cosmic voids, the voids repel the surrounding matter, resulting in an expansion without any dark energy.  The argument demonstrating that matter and anti-matter repel is straightforward, but ultimately incorrect.  The argument is as follows: the CPT theorem guarantees that (flat-space) field theories are invariant under the combination operations of charge conjugation (C), parity (P), and time reversal (T).  Villata postulates that general relativity is also CPT invariant.  To derive how anti-matter responds to matter, he applies a partial CPT transformation to the geodesic equation, applying it only to the ``particle terms,'' and keeping the field terms invariant.  The result is a net sign change in the geodesic equation, indicating an anti-gravity effect.  It is the purpose of this letter to point out two problems with this proposal.  The first problem relates specifically to the behavior of photons, and the second to the use of partial CPT transformations in general.

The geodesic equation in general relativity provides the response of matter (or anti-matter) to a gravitational field and reads in local coordinates
\begin{equation}\label{eq:geo}
\frac{d^2x^\lambda}{d\tau^2}=
\Gamma^{\lambda}_{\;\;\mu\nu}\frac{dx^\mu}{d\tau}\frac{dx^\nu}{d\tau}.
\end{equation}
Here the $x^\mu$ are the spacetime coordinates, $\Gamma$ the Christoffel symbols representing the gravitational field, and $\tau$ an affine parameter of the world line, which can taken as proper time for a massive particle.  The charge conjugation operator C acts trivially on all spacetime quantities.  This leaves parity P and time-reversal T, which change the handedness of space and time, respectively.  The combination PT thus flips all of the spacetime coordinates, PT$:x^\mu\to -x^\mu$.  We note that the affine parameter is invariant under this transformation.  

Following Villata's prescription we apply the CPT transformation to the particles terms (acceleration and velocity) of Eq.\ \ref{eq:geo}, yielding
\begin{align}\label{eq:geo2}
-\frac{d^2x^\lambda}{d\tau^2}&=
\Gamma^{\lambda}_{\;\;\mu\nu}\left(-\frac{dx^\mu}{d\tau}\right)\left(-\frac{dx^\nu}{d\tau}\right)\\
\frac{d^2x^\lambda}{d\tau^2}&=-
\Gamma^{\lambda}_{\;\;\mu\nu}\frac{dx^\mu}{d\tau}\frac{dx^\nu}{d\tau}.
\end{align}
Villata argues that applying this partial CPT transformation represents the response of anti-matter (transformed velocity and acceleration) to a matter-generated (untransformed) gravitational field.

We now voice our first objection: this prescription predicts incorrectly the behavior of photons.  Photons are known to be attracted to matter, as spectacularly demonstrated by gravitational lensing.  However, the above prescription applied to a photon is supposed to yield the behavior of an anti-photon in the same matter-generated gravitational field.  The sign change in the geodesic equation implies that anti-photons are repelled by matter-generated gravitational fields.  But a photon is its own antiparticle, thus yielding the absurdity that a photon is both attracted to and repelled by a gravitating mass.  The same problem will arise for any particle that is its own antiparticle.

We now turn to our second and more general objection, which is the partial application of the CPT transformation to the geodesic equation.  The PT transformation flips the coordinate $x^\mu$ in a coordinate patch about the worldline in question, and thus changes velocities and accelerations into their negatives.  However, changing the coordinates in this way must also change every other quantity expressed in those same coordinates.  In particular the Christoffel symbols, which are the expression of the gravitational field in the local coordinates $x^\mu$, \emph{must} also transform under this coordinate transformation.  The only consistent way to apply the CPT (coordinate) transformation is to apply it to the entire equation.  This introduces a sign change in $\Gamma$, thus showing the equation is invariant under CPT.  Of course, the result could not be otherwise since the geodesic equation is by construction invariant under general coordinate transformations.

This partial CPT transformation is indirectly justified in \cite{V2011} by applying it in a similar way to the Lorentz force equation of electrodynamics, and showing that this gives the correct electromagnetic behavior of anti-matter.  Specifically, there is the (partial) transformation
\begin{equation}
{\rm CPT} :\frac{d^2x^\mu}{d\tau^2}=\frac{q}{m}F^\mu_{\;\;\nu}\frac{dx^\nu}{d\tau}\mapsto
\frac{d^2x^\mu}{d\tau^2}=-\frac{q}{m}F^\mu_{\;\;\nu}\frac{dx^\nu}{d\tau}.
\end{equation}
A net minus is obtained because the acceleration and velocity each pick up a sign from PT, but so does the charge $q$ from C (the Faraday tensor is left invariant).  However, this partial transformation is equally as inappropriate here as in the gravitational case, and for the same reasons.  It just happens to give the ``right answer'' because in the end we have effectively substituted $q\mapsto -q$.  Anti-matter or not, changing the sign of the charge reverses the direction of the Lorentz force.  We obtain the right answer, but for the wrong reason.

While the application of CPT invariance to general relativity to derive anti-gravitational effects between matter and anti-matter is a simple and novel one, ultimately it cannot work.  We have shown that this procedure incorrectly predicts the behavior of photons and that consistency demands applying CPT to the entire geodesic equation, yielding invariance.  We conclude that the observation of anti-gravitational effects in general relativity requires true exotic matter (or energy), and cannot be observed with mundane matter (or anti-matter).

\end{document}